\documentclass[aps,pra,twocolumn,10pt,showpacs,preprintnumbers,amsmath,amssymb,floatfix,longbibliography,superscriptaddress]{revtex4-2}

\usepackage{graphicx}
\usepackage{dcolumn}
\usepackage{bm}
\usepackage{xcolor}
\usepackage{mathrsfs}

\newcommand{\bra}[1]{\langle #1|}
\newcommand{\ket}[1]{|#1\rangle}
\newcommand{\braket}[2]{\langle #1|#2\rangle}

\newcommand{\ITP}{Institute for Theoretical Physics, Institute of Physics, University of Amsterdam,\\Science Park 904, 1098 XH Amsterdam, The Netherlands}
\newcommand{\QS}{QuSoft, Science Park 123, 1098 XG Amsterdam, The Netherlands}
\newcommand{\TU}{Department of Physics and Astronomy, Trinity University, San Antonio, Texas 78212, USA}
\newcommand{\VWI}{Van der Waals-Zeeman Institute, Institute of Physics, University of Amsterdam, 1098 XH Amsterdam, Netherlands}
\newcommand{\NA}{Department of Physics, the United States Naval Academy, Annapolis, Maryland, 21402, USA}
\newcommand{\ITAMP}{ITAMP, Center for Astrophysics, Harvard \& Smithsonian, 60 Garden Street, Cambridge, Massachusetts 02138, USA}
\newcommand{\NNF}{NNF Quantum Computing Programme, Niels Bohr Institute, University of Copenhagen,\\Blegdamsvej 17, 2100 Copenhagen, Denmark}

\definecolor{frenchblue}{rgb}{0.0, 0.45, 0.73}
\definecolor{kellygreen}{rgb}{0.3, 0.73, 0.09}

\usepackage{ulem}
\usepackage{physics}
\usepackage{multirow}

\usepackage{makecell}

\usepackage{hyperref}
\usepackage{xcolor}
\hypersetup{
    colorlinks,
    linkcolor={red!70!black},
    citecolor={green!65!black},
    urlcolor={blue!80!black}
}

\begin{document}
\title{Confinement-induced resonances for the creation of quasi-one-dimensional ultracold gases of alkali–alkaline-earth dimers}
\author{Lorenzo \surname{Oghittu}}
\affiliation{\ITP}
\affiliation{\QS}
\author{Premjith \surname{Thekkeppatt}}
\affiliation{\VWI}
\affiliation{\NNF}
\author{Nirav P. \surname{Mehta}}
\affiliation{\TU}
\affiliation{\ITP}
\affiliation{\ITAMP}
\author{Seth T. \surname{Rittenhouse}}
\affiliation{\NA}
\affiliation{\ITP}
\affiliation{\ITAMP}
\author{Klaasjan \surname{van Druten}}
\affiliation{\VWI}
\affiliation{\QS}
\author{Florian \surname{Schreck}}
\affiliation{\VWI}
\affiliation{\QS}
\author{Arghavan \surname{Safavi-Naini}}
\affiliation{\ITP}
\affiliation{\QS}

\date{\today}

\begin{abstract}
We theoretically investigate the role of confinement-induced resonances (CIRs) in low-dimensional ultracold atomic mixtures in the formation of weakly bound dimers. To this end, we examine the scattering properties of a binary atomic mixture confined by a quasi-one-dimensional (quasi-1D) potential. In this regime, the interspecies two-body interaction is modeled as an effective 1D zero-range pseudopotential, with a coupling strength $g_\mathrm{1D}$ derived as a function of the three-dimensional scattering length $a$. This framework enables the study of CIRs in harmonically confined systems, with particular attention paid to the case of mismatched transverse trapping frequencies of the two atomic species. Finally, we consider the Bose-Fermi mixture of \textsuperscript{87}Rb and \textsuperscript{87}Sr, and  identify values of the experimentally accessible parameters for which CIRs can be exploited to create weakly bound molecules.
\end{abstract}

\maketitle

\section{Introduction}
\begin{figure}[ht!]
    \centering
    \includegraphics[width=\linewidth]{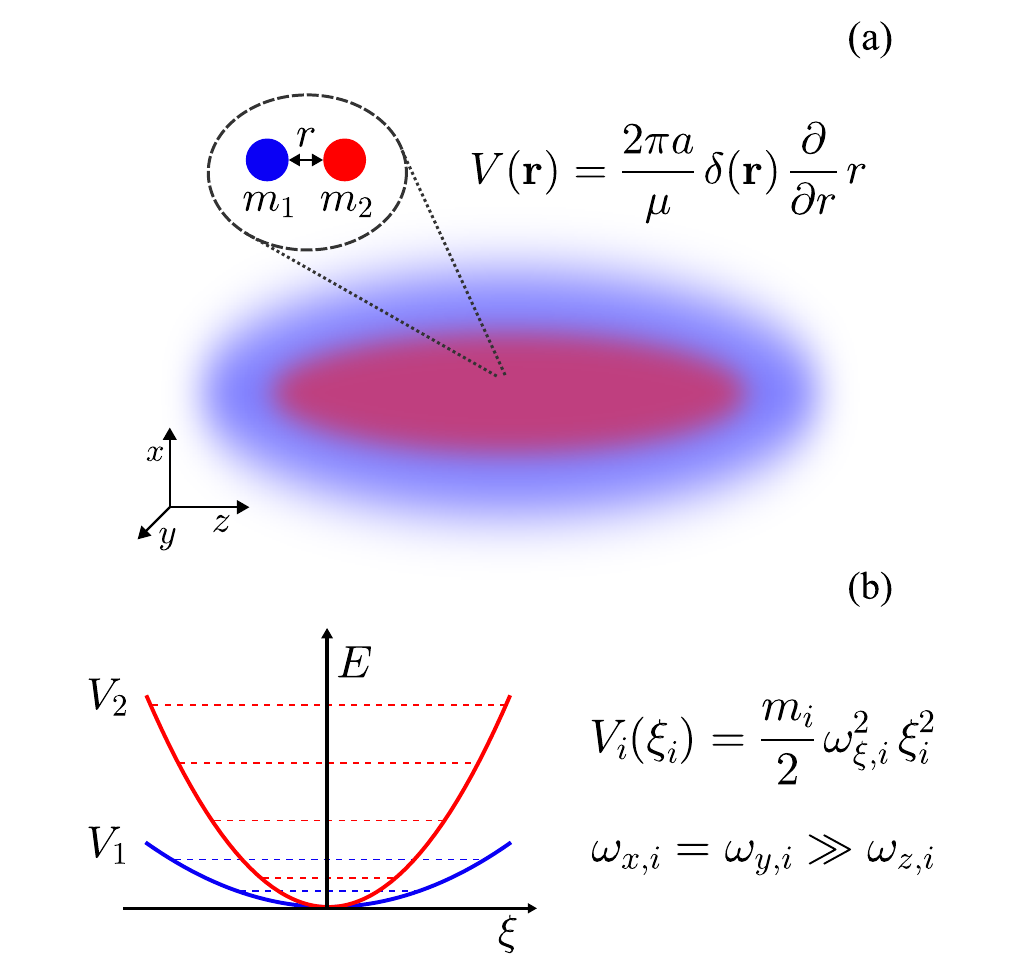}
    \caption{\textbf{(a)} Scheme of the system: a mixture of two species with masses $m_1$ and $m_2$ confined by quasi-one-dimensional potentials with different trapping frequencies. The interspecies interaction is represented by the regularized zero-range pseudopotential $V(r)$, with $r$ being the interparticle distance, $\mu$ the reduced mass and $a$ the 3D scattering length. \textbf{(b)} Harmonic potentials along direction $\xi=x,y,z$ for the two species $i=1,2$. Direction $z$ is the direction of weak confinement.}
    \label{fig:scheme}
\end{figure}

Ultracold gases of polar molecules have emerged as a rapidly advancing field of research~\cite{Langen_NP24}. Owing to their long-range, anisotropic dipole-dipole interactions, these systems offer a highly versatile platform for a wide range of applications, including quantum simulation~\cite{Micheli_NP06}, precision measurement~\cite{Roussy_Sc23,DeMille_NP24}, and quantum chemistry~\cite{Bohn_Sc17}. However, the creation of an ultracold gas of dipolar molecules poses significant challenges.
Experimentally, two methods for creating ultracold gases of ground-state molecules are commonly employed. The first consists of starting with a thermal gas of molecules and cooling it down via laser techniques~\cite{Stuhl_PRL08,Shuman_Nat10,Mitra_Sc20}. Yet, the complicated internal, electronic, rotational, and vibrational structure of the molecules can limit their ability to scatter enough photons, often making laser cooling ineffective for many molecular species~\cite{Jorapur_PRL24}. The second method involves starting from an ultracold gas of unbound atoms and associating them to weakly bound molecules through adiabatic passage across a resonance. These weakly bound molecules can then be directly associated to their rovibrational ground state via stimulated Raman adiabatic passage (STIRAP)~\cite{Ni_Sc08,Danzl_Sc08,Takekoshi_PRL14}. Forming the initial gas of weakly bound molecules usually requires the system to have a readily accessible broad magnetic Feshbach resonance~\cite{Chin_RMP10} as seen in many bialkali systems, e.g. K-Rb~\cite{Ni_Sc08}. Because such broad resonances may not be available in some ultracold mixtures, for example due to a weak magnetic response of the atoms or the resonance appearing at very large magnetic fields~\cite{Ciamei_PRL22}, a different approach becomes necessary. 
In this work, we theoretically investigate the use of a different kind of resonance as a pathway for creating weakly bound molecules: the confinement-induced resonance (CIR)~\cite{Olshanii_PRL98,Bergeman_PRL03,Peano_NJP05,Haller_PRL10,Sala_PRL13,Sala_PRA16,Capecchi_PRL23}. 
CIRs arise in ultracold gases when atoms experience a strong confinement along one or more directions that effectively reduces their dynamics to lower-dimensional regimes. 
Here, we focus on quasi-one-dimensional (quasi-1D) mixtures of alkali and alkaline-earth atoms.
The corresponding heteronuclear dimers constitute a particularly interesting system thanks to their sizable electric and magnetic dipole moments~\cite{Zuchowski_PRA14}. Unlike bi-alkali dimers, however, the formation of these molecules via magneto-association poses considerable challenges. This is primarily due to the non-magnetic ground state of alkaline-earth atoms, which is responsible for exceedingly narrow Feshbach resonances~\cite{Zuchowski_PRL10,Brue_PRA13,Barbe_Nat18}. 
In this context, the use of CIRs emerges as an appealing alternative. Taking into consideration the relevant Bose-Fermi mixture of \textsuperscript{87}Rb and \textsuperscript{87}Sr confined by a quasi-1D harmonic potential (see Fig.~\ref{fig:scheme}), we perform a numerical analysis to characterize the emergence of CIRs. We emphasize that the formation of molecules via two-body CIRs requires a coupling between the center of mass and relative motion, as established in Ref.~\cite{Sala_PRL13}. In the present case, such coupling is introduced by the mismatch in trapping frequencies between the two atomic species, giving rise to narrow inelastic resonances associated with excited center-of-mass bound states. 
In light of this mechanism, our investigation ultimately aims to identify suitable values of tunable parameters that could support ongoing experimental efforts toward realizing ultracold \textsuperscript{87}Rb-\textsuperscript{87}Sr dimers.
\newpage

This paper is organized as follows: 
in Sec.~\ref{sec:two-body_problem} we describe the two-body problem consisting of one particle of each species interacting via a contact potential. 
In Sec.~\ref{sec:scattering_solutions} we show how scattering properties are extracted and we provide explicit expressions for the specific case of harmonic confinement. These two sections closely follow the description in Ref.~\cite{Peano_NJP05}, and are reported here for the sake of completeness. 
In Sec.~\ref{sec:spectrum_1D}, we describe how the spectrum is computed in the 1D limit. 
In Sec.~\ref{sec:analysis}, we present a numerical analysis based on the previous sections. Some comments on how our results are relevant for ongoing experiments are provided in Sec.~\ref{sec:experiments}. Finally, the conclusions and outlook are summarized in Sec.~\ref{sec:conclusions}.

\section{Two-body problem} 
\label{sec:two-body_problem}

\subsection{Hamiltonian and Schr\"{o}dinger equation}
\label{subsec:hamiltonian_and_schrodinger}
We consider two different atomic species with mass $m_1$ and $m_2$. Each of them is confined by a two-dimensional external potential $V_i(\mathbf{x}_{\perp,i})$, where $\mathbf{x}_{\perp,i}=(x_i,y_i)$ indicates the coordinates of species $i$ along the transversal direction of confinement. The longitudinal coordinate is labeled with $z_i$ and the corresponding confinement is assumed to be negligible. Moreover, we assume that we are in the ultracold regime, where only $s$-wave scattering is relevant to describe the atom-atom interaction. The latter can therefore be modeled by the zero-range regularized pseudopotential
\begin{equation}
    V(\mathbf{r})=\frac{2\pi a}{\mu}\delta(\mathbf{r})\frac{\partial}{\partial r}r,
\label{eq:pseudopotential}
\end{equation}
where $\mathbf{r}$ is the separation between the two atoms, $\mu$ is the reduced mass and $a$ is the three-dimensional (3D) intra- or interspecies scattering length. Here and throughout the paper we set $\hbar=1$. The Hamiltonian of a system consisting of one particle of each species is
\begin{equation}
    H=\sum_{i=1,2}\bigg[\frac{\mathbf{p}_i^2}{2m_i}+V_i(\mathbf{x}_{\perp,i})\bigg]+V(|\mathbf{x}_1-\mathbf{x}_2|).
\label{eq:tot_H}
\end{equation}
It is convenient to transform Eq.~\eqref{eq:tot_H} in the center-of-mass and relative coordinates defined by
\begin{equation}
    \mathbf{R}=\frac{m_1\mathbf{x}_1+m_2\mathbf{x}_2}{M},\qquad \mathbf{r}=\mathbf{x}_1-\mathbf{x}_2,
\label{eq:R_and_r}
\end{equation}
with the corresponding momenta
\begin{equation}
    \mathbf{P}=\mathbf{p}_1+\mathbf{p}_2,\qquad \mathbf{p}=\frac{m_2\mathbf{p}_1+m_1\mathbf{p}_2}{M},
\label{eq:P_and_p}
\end{equation}
where $M=m_1+m_2$. We note that the longitudinal center-of-mass coordinate is decoupled from the other degrees of freedom because the confinement is purely transversal. Hence, the system is entirely characterized by the set of coordinates $(\mathbf{R}_\perp,\mathbf{r})$  or, alternatively, by $(\mathbf{x}_{\perp,1},\mathbf{x}_{\perp,2},z)$. Here, $\mathbf{R}_\perp$ is the center-of-mass transversal coordinate, while $z$ is the longitudinal relative coordinate. Analogous notation will be adopted for the momenta. The Hamiltonian can be written as
\begin{equation}
    H=H_\parallel+H_{\perp}^{(1)}+H_{\perp}^{(2)}+V(r),
\label{eq:tot_H_new}
\end{equation}
with $H_\parallel=p_z^2/(2\mu)$ and
\begin{equation}
    H_{\perp}^{(i)}=\frac{p_\perp^2}{2m_i}+V_i(\mathbf{x}_{\perp,i}).
\end{equation}
With the aim of computing the scattering properties of such a system, we start from the two-particles Schr\"{o}dinger equation 
\begin{align}
    (H_0-E)\Psi(\mathbf{R}_\perp,\mathbf{r})&=-V(\mathbf{r})\Psi(\mathbf{R}_\perp,\mathbf{r})\nonumber\\[1ex]
    &=\frac{f(\mathbf{R}_\perp)}{2\mu}\delta(\mathbf{r}),
\label{eq:Scrhoedinger}
\end{align}
where we defined the noninteracting Hamiltonian $H_0=H-V$ and, in the second line, we explicitly took into account the pseudopotential in Eq.~\eqref{eq:pseudopotential} by imposing Bethe-Peierls boundary conditions
\begin{equation}            
    \Psi(\mathbf{R}_\perp,\mathbf{r}\rightarrow0)\simeq\frac{f(\mathbf{R}_\perp)}{4\pi r}\Big(1-\frac{r}{a}\Big).
\label{eq:boundary_conditions}
\end{equation}
The general solution to Eq.~\eqref{eq:Scrhoedinger} can be formally written as
\begin{align}
    \Psi(\mathbf{R}_\perp,\mathbf{r})=&\Psi_0(\mathbf{R}_\perp,\mathbf{r})\nonumber\\[1ex]
    &+\int_{\mathbb{R}^2}d\mathbf{R}_\perp'\,G_E(\mathbf{R}_\perp,\mathbf{r};\mathbf{R}_\perp',0)\frac{f(\mathbf{R}_\perp')}{2\mu},
\label{eq:formal_solution}
\end{align}
where the first term on the right-hand side is the solution to the homogeneous equation $(H_0-E)\Psi_0=0$, while the second is the particular solution written in terms of the Green's function $G_E=(H_0-E)^{-1}$. Therefore, all the information on the scattering between the two particles is contained in $f(\mathbf{R}_\perp)$. Later in the paper, we shall see how to extract it.

Before proceeding, we remark that the energy $E$ indicates the total energy of the system
\begin{equation}
    E=\frac{k^2}{2\mu}+E_{n_1}^{(1)}+E_{n_2}^{(2)},
\end{equation}
where $k$ is the longitudinal relative momentum and $n_i$ is the transverse quantum number of particle $i$. States with fixed $n_1$ and $n_2$ are referred to as channels, which are said to be open when $E\ge E_{n_1}^{(1)}+E_{n_2}^{(2)}$ and closed otherwise.

\subsection{Imaginary-time propagator}
\label{subsec:imaginary-time_propagator}
Let us write the Green's function of the system as
\begin{equation}
    G_E(\mathbf{R}_\perp,\mathbf{r};\mathbf{R}_\perp',0)=\int_0^\infty dt\,e^{Et}G_t(\mathbf{R}_\perp,\mathbf{r};\mathbf{R}_\perp',0),
\label{eq:Greens}
\end{equation}
with the imaginary-time propagator~\cite{Feynman_Hibbs}
\begin{equation}    
G_t(\mathbf{R}_\perp,\mathbf{r};\mathbf{R}_\perp',0)=\bra{\mathbf{R}_\perp,\mathbf{r}}e^{-H_0 t}\ket{\mathbf{R}_\perp',0}.
\label{eq:im-t_propagator}
\end{equation}
According to Eq.~\eqref{eq:tot_H_new}, we can factorize the longitudinal and transversal contribution. The former is the imaginary-time propagator of a free particle
\begin{equation}
    \bra{z}e^{-H_\parallel t}\ket{z'}=\Big(\frac{\mu}{2\pi t}\Big)^{1/2}e^{-(z-z')^2\mu/(2t)},
\label{eq:par_prop}
\end{equation}
while the latter reads
\begin{equation}
    \bra{\mathbf{x}^{}_{\perp,i}}e^{-H_\perp^{(i)} t}\ket{\mathbf{x}_{\perp,i}'}=\sum_n e^{-E_n^{(i)}t}\psi_n^{(i)}(\mathbf{x}_{\perp,i})\Bar{\psi}_n^{(i)}(\mathbf{x}_{\perp,i}'),
\label{eq:perp_prop}
\end{equation}
with $\psi_n^{(i)}$ the eigenstates of $H_\perp^{(i)}$ with eigenvalue $E_{n}^{(i)}$. Hence, we can write Eq.~\eqref{eq:im-t_propagator} as follows
\begin{align}
    G_t(\mathbf{R}_\perp,\mathbf{r};\mathbf{R}_\perp',0)=\sqrt{\frac{\mu}{2\pi t}}e^{-z^2\mu/(2t)}\nonumber\\[1ex]
    \times\prod_{i=1,2}\sum_{n_i}e^{-E^{(i)}_{n_i}t}\psi_{n_i}^{(i)}(\mathbf{x}_{\perp,i}&)\Bar{\psi}_{n_i}^{(i)}(\mathbf{x}_{\perp,i}').
\label{eq:G_t_extended}
\end{align}
We observe from the definition in Eq.~\eqref{eq:im-t_propagator} that the integrand in Eq.~\eqref{eq:Greens} decays with $\mathrm{exp}[-E_Bt]$ at large times, where $E_B=E_0-E$ is the binding energy and $E_0=E_0^{(1)}+E_0^{(2)}$ is the ground state energy of $H_0$. However, the term with $n_1=n_2=0$
gives a contribution to the Green's function proportional to $1/\sqrt{E_B}$~\footnote{This can be verified by simply integrating the $n_1=n_2=0$ in Eq.~\eqref{eq:G_t_extended} according to Eq.~\eqref{eq:Greens}, and by applying the definition of $\zeta_E$ in Eq.~\eqref{eq:Zeta}.}, which is divergent when $E_B$ approaches zero.
Therefore, as soon as the lowest channel opens at $E=E_0$, it is useful to separate its diverging contribution from that of the closed channels. We do that by substituting the transversal Hamiltonian $H_{\perp,i}$ in Eq.~\eqref{eq:perp_prop} with its projection onto the Hilbert subspace orthogonal to the open channel. In the next section, we indicate with a tilde the operators that require such a precaution. Finally, we note that this approach can be generalized to the case where excited states become energetically open.

\section{Scattering solutions} 
\label{sec:scattering_solutions}
The aim of this section is to derive solutions to the two-body problem defined above. 

\subsection{General results}
\label{subsec:general_results}
The incoming state in the two-particle problem is conveniently written as
\begin{align}
    \Psi_0(\mathbf{x}_{\perp,1},\mathbf{x}_{\perp,2},z)&=e^{ikz}\,\psi_0^{(1)}(\mathbf{x}_{\perp,1})\psi_0^{(2)}(\mathbf{x}_{\perp,2})\nonumber\\[1ex]
    &\equiv e^{ikz}\,\psi_0(\mathbf{R}_{\perp},\mathbf{r}_{\perp}),
\end{align}
with $\psi_0^{(i)}$ the single-particle transversal ground state and $k=\sqrt{2\mu(E-E_0)}$ the longitudinal relative momentum. We can write the Green's function as 
\begin{align}
    G_E(\mathbf{R}_\perp,\mathbf{r};\mathbf{R}_\perp',0)=\psi_0&(\mathbf{R}_\perp,\mathbf{r}_{\perp})\Bar{\psi}_0(\mathbf{R}_\perp',0)\frac{i\mu}{k}e^{ik|z|}\nonumber\\[1ex]
    &+\int_0^\infty dt\,e^{Et}\widetilde{G}_t(\mathbf{R}_\perp,\mathbf{r};\mathbf{R}_\perp',0),
\label{eq:G_split}
\end{align}
where we used Eq.~\eqref{eq:Greens} and we explicitly separated the contribution from the open channel. Note that $\widetilde{G}_t$ is simply obtained by subtracting from the sum in Eq.~\eqref{eq:G_t_extended} the open channel contributions, i.e. the term with $n_1=n_2=0$ in our case. In the limit $\mathbf{r}\to 0$, we can once again impose the boundary conditions in Eq.~\eqref{eq:boundary_conditions} and obtain the following integral equation for $f(\mathbf{R}_\perp)$:
\begin{align}
    -\frac{f(\mathbf{R}_\perp)}{4\pi a}=\int_{\mathbb{R}^2}d\mathbf{R}'_\perp\,\widetilde{\zeta}_E(\mathbf{R}_\perp,\mathbf{R}'_\perp)f(\mathbf{R}'_\perp)+\psi_0(\mathbf{R}_\perp,0)\nonumber\\[1ex]
    +\frac{i\psi_0(\mathbf{R}_\perp,0)}{2k}\int_{\mathbb{R}^2}d\mathbf{R}'_\perp\,\Bar{\psi}_0(\mathbf{R}_\perp',0)f(\mathbf{R}'_\perp),
\label{eq:integral_equation}
\end{align}
where we defined the integral kernel
\begin{align}
    \widetilde{\zeta}_E(\mathbf{R}_\perp,\mathbf{R}_\perp')=\lim_{r\to 0}\frac{1}{2\mu}\Big[\widetilde{G}_E&(\mathbf{R}_\perp,\mathbf{r};\mathbf{R}_\perp',0)\nonumber\\[1ex]
    &-\delta(\mathbf{R}_\perp-\mathbf{R}_\perp')\frac{\mu}{2\pi r}\Big],
\label{eq:Zeta}
\end{align}
with $\widetilde{G}_E$ being the closed-channel contribution to the Green's function, i.e. the second line of Eq.~\eqref{eq:G_split}. Note that we can absorb the term proportional to the Dirac delta in the definition of $\widetilde{\zeta}_E$ inside the time integral. We do that by writing
\begin{equation}
    \frac{\mu}{2\pi r}=\int_0^\infty dt\,\Big(\frac{\mu}{2\pi t}\Big)^{3/2}e^{-r^2\mu/(2t)},
\end{equation}
which gives
\begin{align}
    \widetilde{\zeta}_E(\mathbf{R}_{\perp},\mathbf{R}_\perp')=\int_0^{\infty}\frac{dt}{2\mu}&\Big[e^{Et}\widetilde{G}_t(\mathbf{R}_{\perp},0;\mathbf{R}_{\perp}',0)\nonumber\\[1ex]
    &-\Big(\frac{\mu}{2\pi t}\Big)^{3/2}\delta(\mathbf{R}_{\perp}-\mathbf{R}_{\perp}')\Big].
\label{eq:Zeta_2}
\end{align}
We refer to Sec.~\ref{subsec:harmonic_confinement} for the explicit derivation of $\widetilde{G}_t$ in the case of harmonic confinement and to Appendix~\ref{app:matrix_elements} for the computation of the matrix elements of $\widetilde{\zeta}_E$.

In order to understand how $\widetilde{\zeta}_E$ is related to the scattering properties, we now expand Eq.~\eqref{eq:integral_equation} in the basis $\ket{j}$ defined by
\begin{equation}
    \ket{f}=\sum_jf_j\ket{j}, \quad f_j=\int_{\mathbb{R}^2}d\mathbf{R}_\perp\braket{j}{\mathbf{R}_\perp}f(\mathbf{R}_\perp),
\label{eq:basis}
\end{equation}
with the ground state defined by $\braket{\mathbf{R}_\perp}{0}=c\,\psi_0(\mathbf{R}_\perp,0)$, where $c$ is a normalization constant. We can then write Eq.~\eqref{eq:integral_equation} in a compact form:
\begin{equation}
    -\frac{\ket{f}}{4\pi a}=\frac{\ket{0}}{c}+\frac{i}{2k}\frac{\ket{0}}{c^2}\braket{0}{f}+\widetilde{\zeta}_E\ket{f},
\label{eq:integral_equation_compact}
\end{equation}
whose formal solution is given by
\begin{equation}
    \ket{f}=\frac{-1/c}{1-i/(ka_\mathrm{1D})}\Big(\widetilde{\zeta}_E+\frac{1}{4\pi a}\Big)^{-1}\ket{0},
\label{eq:compact_solution}
\end{equation}
with
\begin{equation}
    a_\mathrm{1D}=-\frac{2c^2}{\bra{0}\Big[\widetilde{\zeta}_E+1/(4\pi a)\Big]^{-1}\ket{0}}.
\end{equation}
To check that this can, indeed, be identified with the 1D scattering length, we consider the solution for large separation in the longitudinal direction $z$. In this limit, the contribution from the open channel dominates in Eq.~\eqref{eq:G_split} and Eq.~\eqref{eq:formal_solution} gives the scattering solution
\begin{equation}
    \Psi(\mathbf{R}_\perp,\mathbf{r})=\psi_0(\mathbf{R}_\perp,\mathbf{r}_\perp)\Big[e^{ikz}+f_e(k)e^{ik|z|}\Big],
\label{eq:large_z_solution}
\end{equation}
with the even scattering amplitude
\begin{equation}
    f_e(k)=\frac{i}{2k}\int_{\mathbb{R}^2}d\mathbf{R}'_\perp\Bar{\psi}_0(\mathbf{R}'_\perp,0)f(\mathbf{R}'_\perp).
\label{eq:scattering_amplitude}
\end{equation}
Using Eq.~\eqref{eq:compact_solution} and the definition of the basis in Eq.~\eqref{eq:basis} we get $f_e(k)=-1/(1+ika_\mathrm{1D})$, showing that the term $a_\mathrm{1D}$ assumes the role of the 1D scattering length. We can therefore describe the effective atom-atom interaction with a 1D potential 
\begin{equation}
    V_\mathrm{1D}=g_\mathrm{1D}\delta(z-z'),
    \label{eq:V_1D}
\end{equation}
with interaction strength $g_\mathrm{1D}=-1/(\mu a_\mathrm{1D})$. Indicating by $\lambda_n$ the eigenvalues of $\widetilde{\zeta}_E$ and by $\ket{e_n}$ its eigenvectors, we can write
\begin{equation}
    g_\mathrm{1D}=\frac{1}{2\mu c^2}\sum_n\frac{|\braket{0}{e_n}|^2}{\lambda_n+1/(4\pi a)}.
\label{eq:g1D}
\end{equation}
From this definition, we observe that one resonance appears for each eigenvalue $\lambda_n$, provided that the overlap between the corresponding eigenvector and the ground state of the basis does not vanish.

\subsection{Harmonic confinement}
\label{subsec:harmonic_confinement}
Let us consider the case of harmonic confinement. We indicate with $\omega_{\perp,i}$ the transverse frequency corresponding to particle $i$, and we consider the general case where $\omega_{\perp,1}\neq\omega_{\perp,2}$. The single particle transverse propagator in Eq.~\eqref{eq:perp_prop} can be written as the product of two sums along $x_i$ and $y_i$, with wave functions $\psi_{n_i}^{(i)}$ and energies $E_{n_i}$ given by the eigenstates and eigenenergies of the 1D harmonic oscillator in the corresponding direction. Dropping all the indices for brevity, we have $E_n=\omega_\perp(n+1/2)$ and
\begin{equation}
    \psi_n(x)=\frac{1}{\sqrt{2^nn!}}\Big(\frac{m\omega_\perp}{\pi}\Big)^{\frac{1}{4}}e^{-\frac{m\omega_\perp}{2}x^2}H_n\big[\sqrt{m\omega_\perp}\,x\big],
\label{eq:1D_oscillator_wavefunction}
\end{equation}
where $H_n$ are Hermite polynomials. The contribution from each direction is computed by using the following identity \cite{Gradshteyn_Ryzhik}:
\begin{align}
    \sum_{n=0}^\infty\frac{1}{2^nn!}H_n(x)&H_n(x')\xi^n=\nonumber\\[1ex]
    =\frac{1}{\sqrt{1-\xi^2}}&\,\mathrm{exp}\bigg[\frac{2xx'\xi-(x^2+x'^2)\xi^2}{1-\xi^2}\bigg],
\label{eq:identity1}
\end{align}
where $\xi=\mathrm{exp}(-\omega_\perp t)$ in our case. The product of the two directions gives
\begin{widetext}
\begin{align}
    \bra{\mathbf{x}^{}_{\perp,i}}e^{-H_\perp^{(i)} t}\ket{\mathbf{x}_{\perp,i}'}=&\frac{m_i\omega_{\perp,i}}{\pi}\frac{e^{-\omega_{\perp,i} t}}{1-e^{-2\omega_{\perp,i} t}} \nonumber \\[1ex]
    &\times\mathrm{exp}\bigg[2m_i\omega_{\perp,i}\frac{e^{-\omega_{\perp,i} t}}{1-e^{-2\omega_{\perp,i} t}}\mathbf{x}_{\perp,i}\cdot\mathbf{x}'_{\perp,i}-m_i\omega_{\perp,i}\bigg(\frac{1}{2}+\frac{e^{-2\omega_{\perp,i} t}}{1-e^{-2\omega_{\perp,i} t}}\bigg)\big(\mathbf{x}_{\perp,i}^2+\mathbf{x'}_{\perp,i}^2\big)\bigg].
\label{eq:singleparticle_prop}
\end{align}
\end{widetext}

Now, the transverse propagator for the two particles is obtained as the product of the two single particle propagators. This can be written in a convenient form when $z\to0$. In this limit, the two particles are on top of each other and $\mathbf{x}_{\perp,i}=\mathbf{R}_\perp$, where $\mathbf{R}_\perp$ is the transverse center-of-mass coordinate. 
Using the definitions $\ell_{\perp,i}=1/\sqrt{m_i\omega_{\perp,i}}$,
\begin{align}
\label{eq:some_definitions}
     &\ell_M=\frac{1}{\sqrt{m_1\omega_{\perp,1}+m_2\omega_{\perp,2}}},\qquad \beta=\frac{\ell_M^2}{\ell_{\perp,1}^2},\nonumber\\[2ex]
    &f(t)=\beta\mathrm{coth}(\omega_{\perp,1}t)+(1-\beta)\mathrm{coth}(\omega_{\perp,2}t),\nonumber\\[1ex]
     &g(t)=\beta\frac{1}{\mathrm{sinh}(\omega_{\perp,1}t)}+(1-\beta)\frac{1}{\mathrm{sinh}(\omega_{\perp,2}t)},
\end{align}
the propagator in the transversal direction can be written in the following compact form:
\begin{align}
    \prod_{i=1,2}\bra{\mathbf{x}^{}_{\perp,i}}e^{-H_\perp^{(i)} t}&\ket{\mathbf{x}_{\perp,i}'}=\frac{e^{-\omega_{\perp,1}t}}{1-e^{-2\omega_{\perp,1}t}}\frac{e^{-\omega_{\perp,2}t}}{1-e^{-2\omega_{\perp,2}t}} \nonumber \\[1ex]
        \times\frac{\beta(1-\beta)}{\pi^2\ell_M^4}\mathrm{exp}&\bigg[-f(t)\frac{\mathbf{R}_\perp^2+\mathbf{R}'^2_\perp}{2\ell_M^2}+g(t)\frac{\mathbf{R}_\perp\cdot\mathbf{R}'_\perp}{\ell_M^2}\bigg].
\label{eq:trans_prop_compact}
\end{align}

As we mentioned in Sec.~\ref{subsec:general_results}, when the lowest channel is open, we need to separate its contribution from Eq.~\eqref{eq:trans_prop_compact}. This is done by subtracting the term corresponding to $n_x=n_y=0$ for the two particles.
Combining the previous results and considering the longitudinal propagator in Eq.~\eqref{eq:par_prop} in the limit $z,z'\to0$, we obtain the imaginary-time propagator projected onto the Hilbert space of closed channels:
\begin{align}
    &\widetilde{G}_t(\mathbf{R}_\perp,0;\mathbf{R'}_\perp,0)=\nonumber\\
    &\sqrt{\frac{\mu}{2\pi t}}\frac{\beta(1-\beta)}{\pi^2\ell_M^4}\Bigg\{\frac{\mathrm{exp}\bigg[-f(t)\frac{\mathbf{R}_\perp^2+\mathbf{R}'^2_\perp}{2\ell_M^2}+g(t)\frac{\mathbf{R}_\perp\cdot\mathbf{R}'_\perp}{\ell_M^2}\bigg]}{\big(1-e^{-2\omega_{\perp,1}t}\big)\big(1-e^{-2\omega_{\perp,2}t}\big)}\nonumber\\[1ex]
    &-\mathrm{exp}\bigg[-\frac{\mathbf{R}_\perp^2+\mathbf{R}'^2_\perp}{2\ell_M^2}\bigg]\Bigg\}e^{-(\omega_{\perp,1}+\omega_{\perp,2})t}.
\label{eq:complete_propagator_closed}
\end{align}
The previous equation is used in Eq.~\eqref{eq:Zeta_2} to obtain $\widetilde{\zeta}_E$. The latter is then diagonalized to obtain the 1D interaction strength $g_\mathrm{1D}$, as discussed in the previous section, with the normalization constant given by $c=\sqrt{\pi}\ell_1\ell_2/\ell_M$ in the case of harmonic confinement. The calculation of the matrix elements of $\widetilde{\zeta}_E$ is detailed in the Appendix.

\begin{figure}[ht!]
    \centering
    \includegraphics[width=\linewidth]{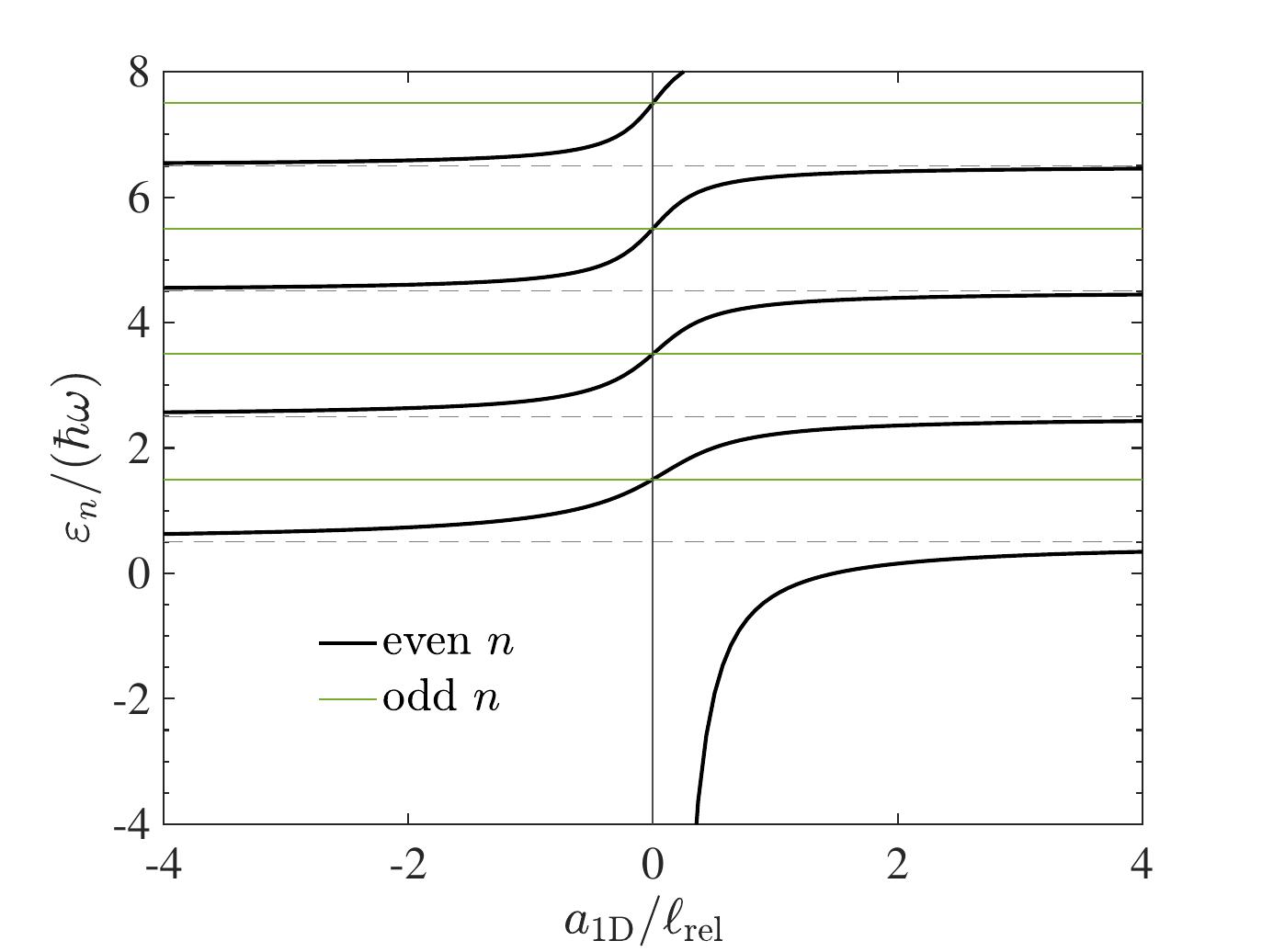}
    \caption{Energy spectrum of $H_\mathrm{rel}$ as a function of the 1D scattering length. Thick black lines: energies for even $n$ approaching the noninteracting values (gray dashed line) at large negative or positive values of $a_\mathrm{1D}$. Thin green lines: energies for odd $n$. Units are rescaled with respect to the energy $\hbar\omega$ and length $\ell_\mathrm{rel}=\sqrt{\hbar/(\mu\omega)}$ of the relative harmonic oscillator.}
    \label{fig:relative_spectrum}
\end{figure}
\section{Spectrum in the 1D limit}
\label{sec:spectrum_1D}
Here, we consider the system described in Sec.~\ref{sec:two-body_problem} in the limit where the transversal trapping is infinitely tight and the particles are forced to move in the longitudinal direction. The longitudinal confinement of the two species, which was previously neglected, is now taken into account and is represented by a harmonic potential with frequencies $\omega_1$ and $\omega_2$. In this scenario, the atom-atom interaction is modeled by a one-dimensional contact potential.

The energy spectrum as a function of the 1D scattering length can be computed similarly to that in Ref.~\cite{Bertelsen_PRA07}. The 1D Hamiltonian is written in terms of the center-of-mass ($Z$,~$P_z$) and relative coordinates ($z$, $p_z$) as
\begin{equation}
    H_\mathrm{1D}=H_\mathrm{com}+H_\mathrm{rel}+\kappa\,zZ.
\label{eq:1D_Hamiltonian}
\end{equation}
The first and second term on the right hand side correspond to the Hamiltonian of the center-of-mass and relative harmonic oscillator:
\begin{align}
    H_\mathrm{com}&=\frac{P_z^2}{2M}+\frac{M\Omega^2}{2}Z^2\nonumber\\
    H_\mathrm{rel}&=\frac{p_z^2}{2\mu}+\frac{\mu\omega^2}{2}z^2+V_\mathrm{1D}(z),
\end{align}
where $V_\mathrm{1D}(z)$ is defined in Eq.~\eqref{eq:V_1D}.
The frequencies are given by
\begin{equation}
    \Omega = \omega_1\sqrt{\frac{1+\lambda}{1+\sigma}},\qquad \omega=\omega_1\sqrt{\frac{\sigma^2+\lambda}{\sigma(1+\sigma)}},
\label{eq:frequencies}
\end{equation}
with $\sigma=m_2/m_1$ and $\lambda=m_2\omega_2^2/(m_1\omega_1^2)$. The last term in Eq.~\eqref{eq:1D_Hamiltonian} is the coupling between the center-of-mass and relative motion with coupling coefficient 
\begin{equation}
    \kappa=m_1\omega_1^2\,\frac{\sigma-\lambda}{1+\lambda}.
\end{equation}
Note that $\omega_1=\omega_2$ implies $\kappa=0$, meaning that the center-of-mass and relative motion separate in that case.
It is convenient to write the matrix elements of $H_\mathrm{1D}$ in the basis given by the product states
\begin{equation}
    \Psi_{N,n}(Z,z)=\Phi_N(Z)\varphi_n(z)
\label{eq:1D_basis}
\end{equation}
with $\Phi_N(Z)$ and $\varphi_n(z)$ being the eigenstates of $H_\mathrm{com}$ and $H_\mathrm{rel}$, respectively. For every $N$, $\Phi_N(Z)$ is simply given by Eq.~\eqref{eq:1D_oscillator_wavefunction} with the appropriate frequency and mass. On the other hand, $\varphi_n(z)$ is given by the same expression only when $n$ is odd, in which case it vanishes at $z=0$ and the contact interaction has no effect. When $n$ is even, the relative wavefunctions are perturbed by the interaction potential and are given by
\begin{equation}
    \varphi_n(x)=A_ne^{-\frac{\mu\omega}{2}z^2}U\bigg(-\frac{\nu_n}{2},\frac{1}{2},\mu\omega z^2\bigg),
\label{eq:relative_wavefunction}
\end{equation}
where $A_n$ is a normalization factor and $U(a,b,z)$ is the Tricomi confluent hypergeometric function. The corresponding eigenvalue is given by $\varepsilon_n=\omega(\nu_n+1/2)$ (see Fig.~\ref{fig:relative_spectrum}), where $\nu_n$ is the non-integer solution to the equation~\cite{Busch_FP98}
\begin{equation}
    \frac{\Gamma\big(-\varepsilon_n/2+1/4\big)}{2\Gamma\big(-\varepsilon_n/2+3/4\big)}=a_\mathrm{1D},
\label{eq:energy_equation_1D}
\end{equation}
with $\Gamma(z)$ being Euler's gamma function and $a_\mathrm{1D}$ the 1D scattering length. Note that the definition of $\nu_n$ can be extended by establishing that $\nu_n=n$ when $n$ is odd. In doing that, the definition in Eq.~\eqref{eq:relative_wavefunction} can be used as a general definition for the relative wavefunction, since it retrieves the wavefunction of the unperturbed harmonic oscillator when $\nu_n$ is a positive integer~\cite{Oldham_Myland_Spanier}. The matrix elements of $H_\mathrm{1D}$ are therefore given by
\begin{align}
    \big[H_\mathrm{1D}\big]_{i,j}=&\Omega\bigg(N_i+\frac{1}{2}\bigg)\delta_{i,j}+\omega\bigg(\nu_{n_i}+\frac{1}{2}\bigg)\delta_{i,j} \nonumber \\[1ex]
    &+\kappa\bra{\varphi_{n_i}(z)}z\ket{\varphi_{n_j}(z)}\bra{\Phi_{N_i}(Z)}Z\ket{\Phi_{N_j}(Z)},
\label{eq:matrix_elements_H1D}
\end{align}
and the energy spectrum is obtained via numerical diagonalization. 
\begin{figure}[ht!]
    \centering
    \includegraphics[width=\linewidth]{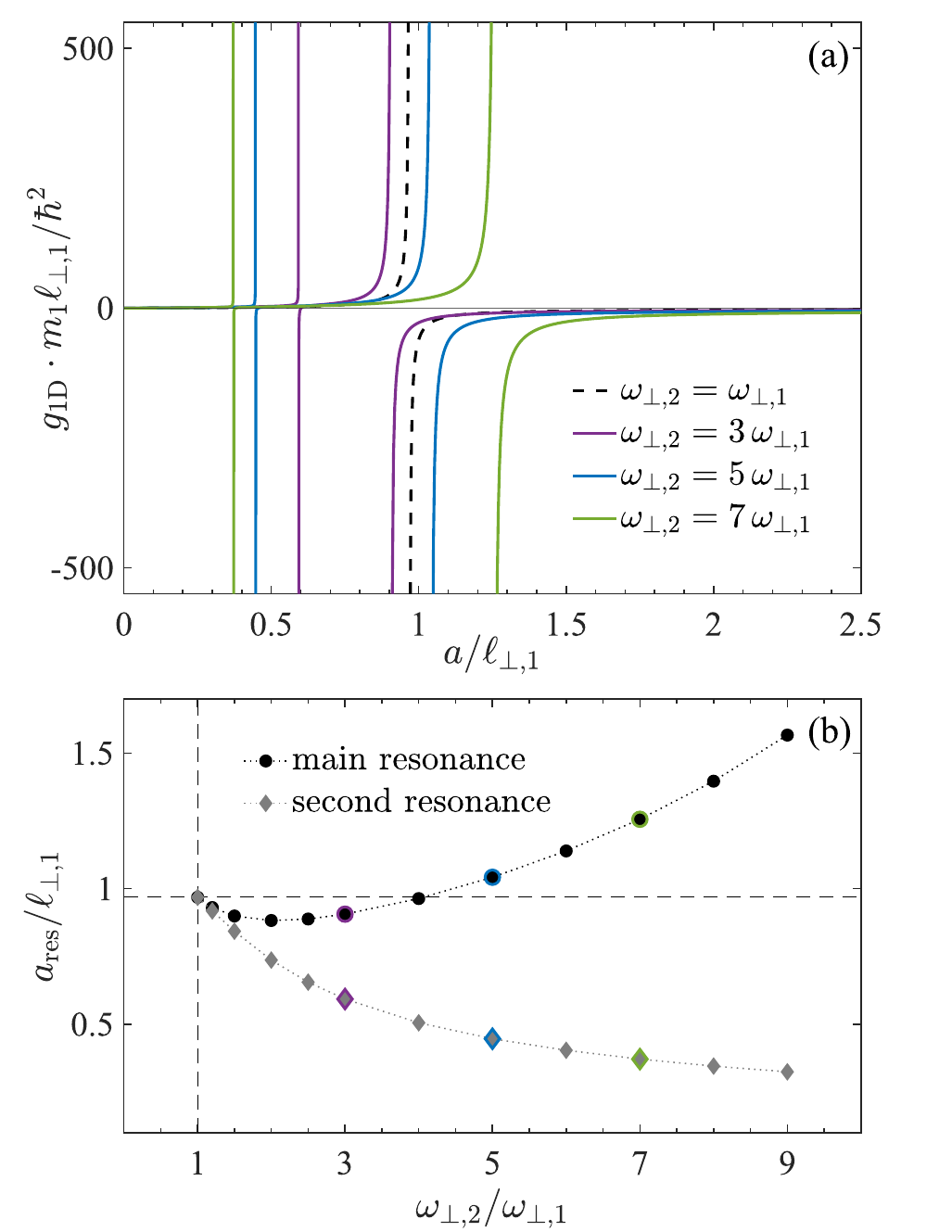}
    \hypertarget{fig:g1D}{}
    \caption{Confinement induced resonances for different frequency ratios. \textbf{(a)} Interaction strength $g_\mathrm{1D}$ as a function of scattering length. \textbf{(b)} Values $a_\mathrm{res}$ of scattering length $a$ for which the main (black circles) and second resonance (gray diamonds) appear. Dashed lines indicate the analytical value obtained for $\omega_{\perp,2}=\omega_{\perp,1}$. The position of the resonances shown in (a) are highlighted with the corresponding color. Dotted lines are only a guide to the eye.}
    \label{fig:g1D}
\end{figure}
\section{Numerical Analysis}
\label{sec:analysis}
The following numerical analysis is devoted to the case of harmonic confinement. Moreover, we focus on the ultracold regime where $E\rightarrow E_0$ and only the ground state is energetically open. Results are therefore based on the derivations in Secs.~\ref{subsec:harmonic_confinement} and~\ref{sec:spectrum_1D}. 

Without loss of generality, we always consider the masses of the two species to be those of \textsuperscript{87}Rb and \textsuperscript{87}Sr, where $m_\mathrm{Rb}\simeq86.9092\,\mathrm{m_u}$ and $m_\mathrm{Sr}\simeq86.9089\,\mathrm{m_u}$ ($\mathrm{m_u}\simeq1.6605\times10^{-27}\,\mathrm{kg}$). Note that, from this point forward, we will use a subscript $1$ to denote quantities related to the strontium atom and a subscript $2$ for those related to the rubidium atom.

\subsection{Resonance position}
Let us start by studying the position of confinement-induced resonances for different values of the ratio $\omega_{\perp,2}/\omega_{\perp,1}$ in the quasi-1D regime. We recall that in this regime the confinement is purely transversal.
In Figure~\ref{fig:g1D}(\hyperlink{fig:g1D}{a}), the interaction strength $g_\mathrm{1D}$ as defined in Eq.~\eqref{eq:g1D} is plotted as a function of the 3D scattering length $a$ for different values of the trap frequency ratio $\omega_{\perp,2}/\omega_{\perp,1}$. 
We observe that in the case where $\omega_{\perp,2}=\omega_{\perp,1}$ (black dashed line), only one resonance appears. This can be interpreted as a zero-energy Feshbach resonance due to the binding energy of the first excited relative motion state 
matching the continuum threshold of the open channel \cite{Bergeman_PRL03,Peano_NJP05}. The position of this resonance is computed analytically in Ref.~\cite{Olshanii_PRL98}. 

As the frequency ratio deviates from unity, an additional narrow resonance appears for lower values of $a$. Figure~\ref{fig:g1D}(\hyperlink{fig:g1D}{b}) shows the values $a_\mathrm{res}$ of $a$ at which the main and second resonances in $g_\mathrm{1D}$ are located for different ratios. 
We observe that the position of the second resonance (gray diamonds) decreases as $\omega_{\perp,2}/\omega_{\perp,1}$ increases. 
Like in the case with equal traps, this resonance has a Feshbach-like interpretation. In this case, however, it is related to the first excited center-of-mass bound state.
Note that, in principle, a distinct resonance is associated with each of the higher excited center-of-mass states. However, their width is vanishing in the range of values that we consider for the trap ratio. 
The position of the main resonance (black circles), on the other hand, presents a minimum around $\omega_{\perp,2}=2\,\omega_{\perp_1}$ and does not have a similar interpretation. We can therefore speculate that the state responsible for its presence is a nontrivial combination of center-of-mass and relative bound states.
\begin{figure}[ht!]
    \centering
    \includegraphics[width=\linewidth]{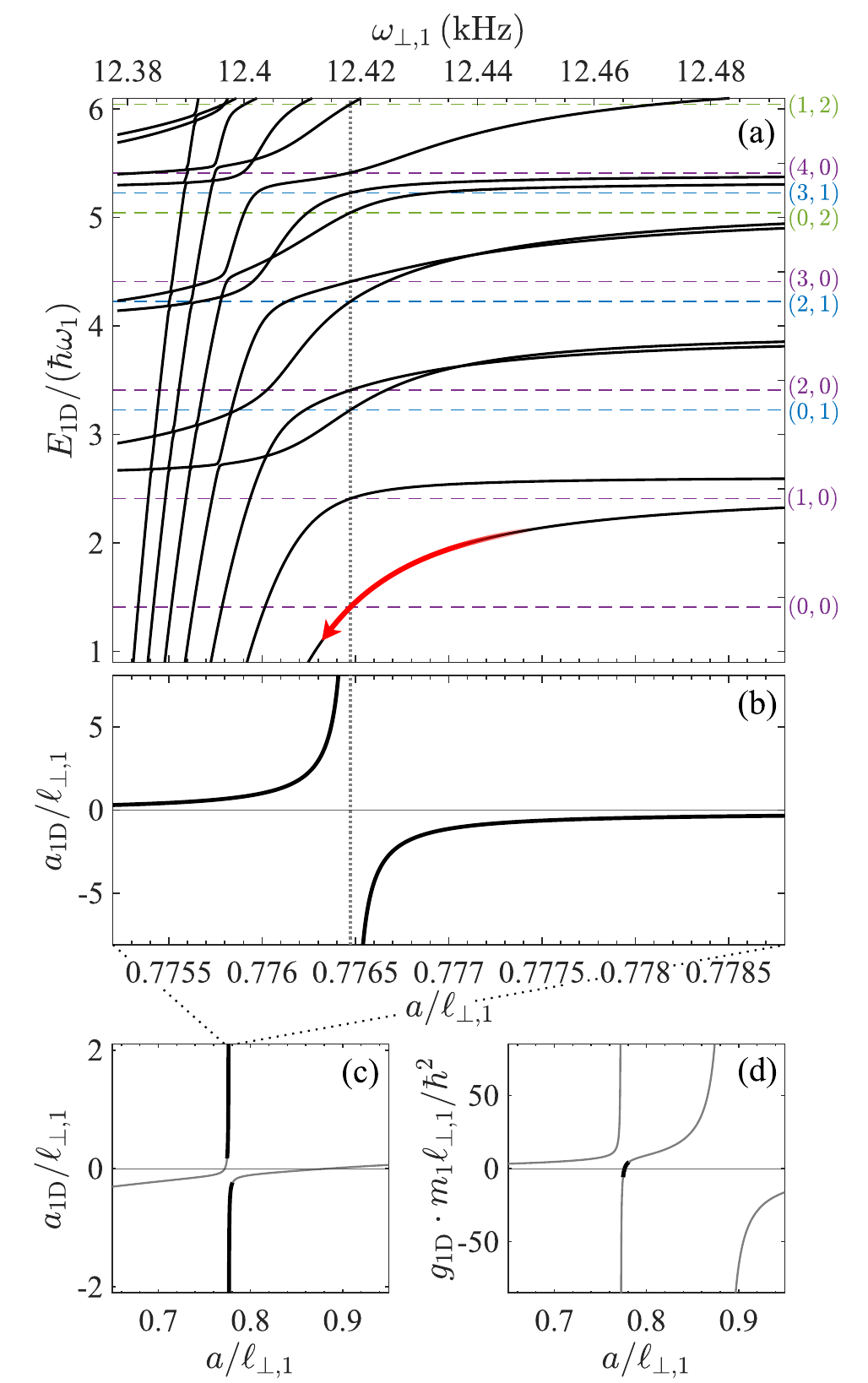}
    \hypertarget{fig:RbSr_1D_spectrum}{}
    \caption{Results for the \textsuperscript{87}Rb-\textsuperscript{87}Sr mixture. \textbf{(a)} Spectrum in the 1D limit. The upper axis indicates the transverse strontium frequency $\omega_{\perp,1}$ considering the fixed interspecies scattering length $a_\mathrm{Sr-Rb}=1420\,a_0$ ($a_0$ the Bohr radius). Horizontal dashed lines correspond to the asymptotic noninteracting states for different values of $(n_1,n_2)$ as indicated on the right of the plot, with $n_1$ and $n_2$ the principal quantum numbers of the longitudinal harmonic oscillator of strontium and rubidium, respectively. Purple lines are for $n_2=0$, blue lines for $n_2=1$ and green lines for $n_2=2$. The vertical dotted line indicates the position of the resonance of $a_\mathrm{1D}$ appearing at $\omega_{\perp,1}=12.418\,\mathrm{kHz}$ or $a=0.77647\,\ell_{\perp,1}$ ($\ell_{\perp,1}$ is the length associated with $\omega_{\perp,1}$). The red arrow indicates a possible pathway for dimer formation: the system is prepared on the right side of the resonance and adiabatically ramped across it to populate a bound state. \textbf{(b)} 1D scattering length $a_\mathrm{1D}=-1/(\mu g_\mathrm{1D})$ in units of $\ell_{\perp,1}$. The bottom axis indicates the 3D scattering length in units of $\ell_{\perp,1}$. \textbf{(c)} 1D scattering length on a broader range. \textbf{(d)} 1D interaction strength $g_\mathrm{1D}$. The values used in previous plots are highlighted in black.}
    \label{fig:RbSr_1D_spectrum}
\end{figure}
\subsection{Spectrum of the \texorpdfstring{\textsuperscript{87}Rb}{Rb}-\texorpdfstring{\textsuperscript{87}Sr}{Sr} mixture}
We now consider the specific case where \textsuperscript{87}Rb and \textsuperscript{87}Sr are confined by the same optical dipole potential.
Due to their different polarizabilities, the two species experience distinct trapping frequencies. Specifically, at the commonly used optical trap wavelength of $1064\,\mathrm{nm}$, the resulting trap frequency ratio is $\omega_{(\perp),2}\simeq1.82\,\omega_{(\perp),1}$ for both the transverse and longitudinal directions. Combining the results derived in Secs.~\ref{sec:scattering_solutions} and \ref{sec:spectrum_1D}, we compute the 1D spectrum as a function of the 3D scattering length $a$. This is done by mapping $a$ to $a_\mathrm{1D}$ using Eq.~(\ref{eq:g1D}) and diagonalizing the Hamiltonian in Eq.~(\ref{eq:matrix_elements_H1D}). 

In Figs~\ref{fig:RbSr_1D_spectrum}(\hyperlink{fig:RbSr_1D_spectrum}{a}) and \ref{fig:RbSr_1D_spectrum}(\hyperlink{fig:RbSr_1D_spectrum}{b}), we present the spectrum and the corresponding values of the 1D scattering length, respectively. They are plotted as a function of the ratio between $a$ and the oscillator length associated with the perpendicular confinement of strontium atoms $\ell_{\perp,1}$, as shown in the lower axis of Figs~\ref{fig:RbSr_1D_spectrum}(\hyperlink{fig:RbSr_1D_spectrum}{b}). The upper axis of panel Figs~\ref{fig:RbSr_1D_spectrum}(\hyperlink{fig:RbSr_1D_spectrum}{a}) indicates the corresponding dependence on the frequency $\omega_{\perp,1}$, which is accessible experimentally and is computed from the ratio $a/\ell_{\perp,1}$ by considering the scattering length $a_\mathrm{Rb-Sr}=1420\,a_0$ between \textsuperscript{87}Rb and \textsuperscript{87}Sr in units of Bohr radius $a_0$. We observe that the avoided crossings between trapped states and excited bound states are found in a narrow range of values around $\omega_{\perp,1}=12.418\,\mathrm{kHz}$ (dotted vertical line) where $a_\mathrm{1D}$ has a pole and the spectrum crosses the asymptotic noninteracting states (dashed horizontal lines). To better understand these features, we recall that the scattering length in the 1D limit is inversely proportional to the interaction strength. Hence, large variations of $a_\mathrm{1D}$ are obtained with small variations of $g_\mathrm{1D}$ around zero. From Figs.~\ref{fig:RbSr_1D_spectrum}(\hyperlink{fig:RbSr_1D_spectrum}{c}) and \ref{fig:RbSr_1D_spectrum}(\hyperlink{fig:RbSr_1D_spectrum}{d}), it is evident that this can be achieved only in the region between the two poles in $g_\mathrm{1D}$, where the latter crosses the horizontal axes and $a_\mathrm{1D}$ presents a narrow resonance. Figure~\ref{fig:relative_spectrum} shows that for large positive values of $a_{\mathrm{1D}}$, the relative Hamiltonian supports a shallow molecular bound state. We propose that such a bound state may be populated by  adiabatically ramping the system toward smaller values of $\omega_{\perp,1}$, thus transferring atoms into the molecular bound state. A possible pathway is indicated by the red arrow in Fig.~\ref{fig:RbSr_1D_spectrum}(\hyperlink{fig:RbSr_1D_spectrum}{a}). The transfer efficiency is expected to be high when the ramping timescales are long compared to $1/\omega_{1}$, such that the system is allowed to adiabatically follow the ground state in Fig.~\ref{fig:RbSr_1D_spectrum}(\hyperlink{fig:RbSr_1D_spectrum}{a}). It is worth noting that additional crossings between bound states and trapped states occur in the range where $a/\ell_{\perp,1}\lesssim5$, i.e. $\omega_{\perp,1}\lesssim700\,\mathrm{kHz}$, which is much more challenging to access in current experiments. Furthermore, it is important to recognize that the reported resonance position is found by assuming $a_\mathrm{Rb-Sr}=1420\,a_0$. Experimental realizations should account for the uncertainty in $a_\mathrm{Rb-Sr}$ and its impact on this estimate.

We finally remark that these results are relevant for ongoing experiments involving the creation of weakly bound \textsuperscript{87}Rb-\textsuperscript{87}Sr dimers. In particular, the spectrum in Fig.~\ref{fig:RbSr_1D_spectrum}(\hyperlink{fig:RbSr_1D_spectrum}{a}) reveals the coupling between excited bound states and lower-lying trapped states. CIRs can therefore be exploited to adiabatically ramp the system from a regime where the two species are in their corresponding unbound trapped state to the one where bound states are occupied. 
We refer to the next section for a comment on this matter.

\section{Relation to Experiments}
\label{sec:experiments}
In this section, we examine the experimental conditions necessary for observing confinement-induced resonances in \textsuperscript{87}Rb–\textsuperscript{87}Sr mixtures and explore their potential as a pathway for the formation of weakly bound dimers.

First, we remark that the observation of CIRs requires a near-threshold molecular bound state with a binding energy comparable to the harmonic trapping frequency. The selection of the \textsuperscript{87}Rb–\textsuperscript{87}Sr isotopologue is motivated by the existence of such a bound state, with a binding energy of approximately $20\,\mathrm{kHz}$, well matched to typical trap frequencies achievable in state-of-the-art optical dipole traps and optical lattices \cite{Alessio_PCCP18}. Tuning the harmonic trapping frequency across a CIR is particularly important in ultracold mixtures with narrow Feshbach resonances, as it enables controlled coupling between the near-threshold bound state and the trapped atomic states.

This requirement narrows the range of suitable experimental platforms for studying CIRs in quasi-one-dimensional regimes. Two main approaches are particularly relevant: the first employs a two-dimensional optical lattice to create tightly confining, tube-like dipole potentials, effectively restricting motion to one dimension. The second consists of using optical tweezers for precise control of individual atoms or small ensembles in highly anisotropic traps. Both methods offer excellent tunability of dimensionality and interaction parameters, providing ideal conditions for observing CIRs.

However, in the \textsuperscript{87}Rb–\textsuperscript{87}Sr system, the near-threshold bound state results in a large three-dimensional scattering length of approximately $1420\,a_0$ leading to strong interspecies interactions and rapid three-body recombination losses. Sympathetic cooling in overlapping traps is therefore inefficient. To address this limitation, the atomic species undergo independent evaporative cooling in spatially separated optical dipole traps before being adiabatically merged into a common quasi-one-dimensional confinement potential. This sequential cooling strategy minimizes interspecies collisions during critical evaporative stages while preserving phase-space density.

The detection of CIRs can be achieved by monitoring atomic loss rates due to enhanced three-body recombination near resonance, although this method becomes less sensitive in quasi-1D systems due to the suppression of three-body recombination in one dimension \cite{mehta2007three}. More precise detection involves measuring interspecies thermalization rates, which increase near a CIR as the effective one-dimensional scattering length diverges \cite{CHO_PRA13}. Alternatively, photoassociation spectroscopy can be employed, where resonantly enhanced atom–molecule coupling leads to increased photoassociation rates near resonance.

Based on these considerations, confinement-induced resonances are expected to provide a promising route toward the formation of weakly bound \textsuperscript{87}Rb–\textsuperscript{87}Sr dimers.

\section{Conclusions}
\label{sec:conclusions}
We studied the emergence of confinement-induced resonances in ultracold mixtures of two atomic species trapped by harmonic quasi-1D potentials with mismatched trapping frequencies. 
We systematically analyzed the behavior of CIRs and their dependence on the ratio between the transverse trapping frequencies $\omega_{\perp,2}/\omega_{\perp,1}$ of the two species (Fig.~\ref{fig:g1D}).
In the case of equal frequencies, it is a well known result (see Ref.~\cite{Olshanii_PRL98}) that a single resonance appears.
Our results show that an additional narrow resonance attached to the first excited center-of-mass bound state emerges at smaller values of the scattering length when the frequencies are different.
Resonances attached to higher center-of-mass excited states are vanishingly narrow for the considered values of the frequency ratio.
Furthermore, we focused on the case where \textsuperscript{87}Rb and \textsuperscript{87}Sr are confined in the same trap, resulting in $\omega_{(\perp),2}/\omega_{(\perp),1}\simeq1.82$ along all directions. The energy spectrum as a function of the transverse frequency of strontium atoms suggests that \textsuperscript{87}Rb-\textsuperscript{87}Sr dimers can be created by properly ramping the system through a resonance that appears in the 1D scattering length at $\omega_{\perp,1}=12.418\,\mathrm{kHz}$ (Fig.~\ref{fig:RbSr_1D_spectrum}).

This work provides a theoretical framework for the efficient formation of ultracold gases of \textsuperscript{87}Rb-\textsuperscript{87}Sr dimers. In doing so, it paves the way for extending the approach to other alkali – alkaline-earth atomic mixtures. Additionally, possible future extensions involve developing an analogous analysis for quasi-two-dimensional geometries.

\section*{Acknowledgements}
A.S.N. was supported by the Dutch Research Council (NWO/OCW) as a part of the Quantum Software Consortium (Project No. 024.003.037), Quantum Delta NL (Project No. NGF.1582.22.030) and ENW-XL grant (Project No. OCENW.XL21.XL21.122). N.P.M. and S.T.R. acknowledge support from the National Science Foundation through ITAMP and in part through NSF Grant No. 2409110, and from Quantum Delta NL through the University of Amsterdam. PT was supported by the Dutch Research Council (NWO) as part of  Quantum Simulation 2.0 (Grant No. 680.92.18.05) and by the Novo Nordisk Foundation (Grant No. NNF23OC0086815). 

\appendix

\section{Matrix elements of \texorpdfstring{$\boldsymbol{\widetilde{\zeta}_E}$}{zeta} for harmonic confinement} 
\label{app:matrix_elements}
Here, we calculate the matrix elements of the operator $\widetilde{\zeta}_E$ as defined in Eq.~\eqref{eq:Zeta_2}. To do that, we consider the basis 
\begin{equation}
    \braket{\mathbf{R}_\perp}{n_x,n_y}=\frac{1}{\ell_M}\psi_{n_x}\bigg(\frac{R_\perp^{(x)}}{\ell_M}\bigg)\psi_{n_y}\bigg(\frac{R_\perp^{(y)}}{\ell_M}\bigg),
\end{equation}
where $\psi_n$ indicates the 1D harmonic oscillator wavefunction in Eq.~\eqref{eq:1D_oscillator_wavefunction}, $R_\perp^{(x)}$ is the first component of the center-of-mass coordinate and $\ell_M$ is defined in Eq.~\eqref{eq:some_definitions}.

We start from the matrix elements of $\widetilde{G}_t$:
\begin{align}
    \big[\widetilde{G}(t)\big]_{\mathbf{n},\mathbf{m}}\equiv&\bra{n_x,n_y}\widetilde{G}_t(\mathbf{R}_\perp,0;\mathbf{R'}_\perp,0)\ket{m_x,m_y}\nonumber\\
    =&\sqrt{\frac{\mu}{2\pi t}}\frac{\beta(1-\beta)}{\pi^2\ell_M^2}e^{-(\omega_{\perp,1}+\omega_{\perp,2})t}\nonumber\\
    \times\bigg\{&\frac{\big[\widetilde{F}(t)\big]_{n_x,m_x}\big[\widetilde{F}(t)\big]_{n_y,m_y}}{(1-e^{-2\omega_{\perp,1}t})(1-e^{-2\omega_{\perp,2}t})}-\pi\delta_{\mathbf{n},0}\delta_{\mathbf{m},0}\bigg\},
\label{eq:G_nm}
\end{align}
where we defined
\begin{align}
    \big[\widetilde{F}(t)\big]_{n,m}=&\int_{-\infty}^\infty dX\int_{-\infty}^\infty dX'\,\Bar{\psi}_n(X)\psi_m(X') \nonumber \\
    &\times\mathrm{exp}\bigg[-\frac{f(t)}{2}(X^2+X'^2)+g(t)XX'\bigg]\nonumber\\
    =&\int_{-\infty}^\infty dX\int_{-\infty}^\infty dX'\,\frac{H_n(X)H_m(X')}{\sqrt{\pi2^{n+m}n!m!}}\nonumber\\
    &\times\mathrm{exp}\bigg[-\frac{1+f(t)}{2}(X^2+X'^2)+g(t)XX'\bigg],
\label{eq:F_nm}
\end{align}
and the rescaled coordinate $X=R_\perp^{(x)}/\ell_M$. Definitions along $y$ are analogous. Note that in the second equality, we explicitly used the 1D harmonic oscillator eigenstates. The integral over $X$ is computed by means of the identity~\cite{Gradshteyn_Ryzhik}
\begin{equation}
    \int_{-\infty}^\infty dz\,e^{-(z-z')^2}H_n(\alpha z)=\sqrt{\pi}(1-\alpha^2)^\frac{n}{2}H_n\bigg[\frac{\alpha z'}{\sqrt{1-\alpha^2}}\bigg],
\end{equation}
with $\alpha(t)=[2/(1+f(t))]^{1/2}$, $z=X/\alpha(t)$ and $z'=g(t)\alpha(t)X'/2$. Completing the square in the exponential in Eq.~\eqref{eq:F_nm}, we are left with
\begin{align}
    \big[\widetilde{F}(t)\big]_{n,m}&=\frac{\alpha\big(1-\alpha^2\big)^{\frac{n}{2}}}{\sqrt{2^{n+m}n!m!}}\int_{-\infty}^\infty dX'\,H_m(X') \nonumber \\
        \times H_n&\bigg(\frac{g\alpha^2}{2\sqrt{1-\alpha^2}}X'\bigg)
        \mathrm{exp}\bigg[-X'^2\bigg(\frac{1}{\alpha^2}-\frac{g^2\alpha^2}{4}\bigg)\bigg],
\label{eq:F_nm_2}
\end{align}
which can be computed using the following identity~\cite{Lord_LMS49}:
\begin{align}
    \int_{-\infty}^\infty dy\,e^{-y^2}H_n(py)H_m(qy)=2^{2M}\Gamma\bigg(M+\frac{1}{2}\bigg)\gamma^n\delta^m \nonumber \\
        \times _2F_1\bigg(-n,-m;-M+\frac{1}{2};\frac{\gamma\delta-pq}{2\gamma\delta}\bigg),
\end{align}
with $M=m+n$, $\gamma^2=p^2-1$, $\delta^2=q^2-1$ and $_2F_1$ the hypergeometric function. To use this identity in Eq.~\eqref{eq:F_nm_2}, we define $y = X'(1/\alpha^2-g^2\alpha^2/4)^{1/2}$ and we perform the change of variable $X'\to y$. Finally, we can write the expression for the matrix element of $\widetilde{\zeta}_E$:
\begin{equation}
    \begin{split}
        \big[\widetilde{\zeta}_E\big]_{\mathbf{n},\mathbf{m}}=\frac{1}{2\mu}\int_0^\infty dt\,\bigg\{e^{Et}\big[\widetilde{G}_t\big]_{\mathbf{n},\mathbf{m}}-\bigg(\frac{\mu}{2\pi t}\bigg)^{\frac{3}{2}}\delta_{\mathbf{n},\mathbf{m}}\bigg\}.
    \end{split}
\end{equation}

\bibliography{bibliography}

\end{document}